\documentclass[12pt]{article}
\usepackage[utf8]{inputenc}
\usepackage{amsmath}
\usepackage{amsthm}
\usepackage{amsfonts}

\usepackage{algorithm}
\usepackage[noend]{algpseudocode}
\usepackage{color}
\usepackage{lineno}
\usepackage{graphicx}
\usepackage{hyperref}
\usepackage{subcaption,natbib}
\usepackage[margin=1in]{geometry}
\usepackage[doublespacing]{setspace}
\usepackage{soul} %%% REMOVE THIS LATER

\title{Bayesian Inference for Polycrystalline Materials}
\author{James Matuk$^1$, Oksana Chkrebtii$^1$, Stephen Niezgoda$^{2,3}$\\
$^1$Department of Statistics, The Ohio State University\\
$^2$Department of Materials Science and Engineering, The Ohio State University \\
$^3$Department of Mechanical and Aerospace Engineering, The Ohio State University}
\date{}

\begin{document}

\maketitle 

\linespread{1}
\begin{abstract}
Polycrystalline materials, such as metals, are comprised of heterogeneously oriented crystals. Observed crystal orientations are modelled as a sample from an orientation distribution function (ODF), which determines a variety of material properties and is therefore of great interest to practitioners. Observations consist of quaternions, 4-dimensional unit vectors reflecting both orientation and rotation of a single crystal. Thus, an ODF must account for known crystal symmetries as well as satisfy the unit length constraint. A popular method for estimating ODFs non-parametrically is symmetrized kernel density estimation. However, disadvantages of this approach include difficulty in interpreting results quantitatively, as well as in quantifying uncertainty in the ODF. We propose to use a mixture of symmetric Bingham distributions as a flexible parametric ODF model, inferring the number of mixture components, the mixture weights, and scale and location parameters based on crystal orientation data. Furthermore, our Bayesian approach allows for structured uncertainty quantification of the parameters of interest. We discuss details of the sampling methodology and conclude with analyses of various orientation datasets, interpretations of parameters of interest, and comparison with kernel density estimation methods.
\end{abstract}

\newpage
\linespread{1.5}

\section{Introduction}

Virtually all engineered and natural inorganic materials are polycrystals \citep{kocks1998texture}. Metals and their alloys, rocks and minerals, as well as advanced engineering and common household ceramics are crystalline in nature. Additionally many common polymers and plastics, such as polyethelyne and polypropylene, are semi-crystalline \citep{flory1978molecular} and polycrystalline biological materials (e.g. hydroxyapetite crystals in bones and teeth) are also not uncommon \citep{ma2009hydroxyapatite}. These materials are comprised of numerous individual crystallites that are heterogeneously oriented, i.e. rotated with respect to a set of reference coordinate axes. The individual crystals exhibit anisotropy, meaning that properties, such as stiffness, strength, conductivity, and permeabilities, vary with crystallographic directions \citep{nye}.  Consequently, the distribution orientations of the individual crystallites is critical to understanding the macroscale response of a polycrystalline material to an applied stimulus. The distribution of orientations, or the rotation of the individual crystallites relative to a fixed measurement reference coordinate frame, is referred to as the crystallographic fabric in geology or texture in materials science and characterized via the orientation distribution function (ODF) \citep{bunge2013texture}.

The set of possible orientations of a crystallite is described by three-dimensional rotation group, $SO(3)$, modulo a set of symmetry elements \citep{bunge2013texture}. The orientation of an individual crystallite can be represented in a number of ways ,e.g. Euler angles, rotation matrices, but for this work, representation by unit-length quaternions is the most convenient. There are two sets of symmetry elements that must be accounted for. Crystal symmetry reflects the inherent rotational symmetry of the crystal \citep{nye}. Specimen, sample, or statistical symmetry reflects symmetry of the sample that is maintained during processing operations, such as uniaxial tension of cylindrical sample or rolling reduction of a rectangular plate \citep{bunge2013texture}. This application and previous methods are described in \cite{niezgoda2013unsupervised} and \cite{niezgoda_2016} among others, where the symmetric Bingham distribution is proposed as an appropriate model for ODFs since it is able to account for the structure of the data. 

In this work, we extend these ideas by modeling the ODF as a finite mixture of symmetric Bingham distributions with an unknown number of components. The mixture distribution framework allows for greater flexibility, and includes the single-component symmetric Bingham distribution as a special case. In contrast with previous work, we utilize a Bayesian approach which allows for structured uncertainty quantification. Due to the variable dimension of the parameter space, a reversible-jump Markov chain Monte Carlo algorithm \citep{green_1995} is employed for posterior sampling. We test our model on a variety of experimentally obtained orientation datasets, and provide interpretations of parameters of interest along with a comparison to kernel density estimation methods, which are commonly used in this area.

\section{Model Formulation}

Let $\mathbf{g} = (g_1,\ldots,g_n)$ denote independent, identically distributed data on the unit sphere embedded in 4-dimensional Euclidean space that we will use to inform our model. An individual quaternion, $g\in \mathcal{S}^3$, represents a relative rotation of a three-dimensional object to reference axes describing orientations of crystals comprising a material. There has been success in the literature in modeling these data using a Bingham distribution,  as in \citet{fallaize_2014}. However, this model may only be adequate if the distribution of the orientations does not contain symmetries, which are a feature of the ODFs of polycrystalline materials of interest.

\cite{niezgoda_2016} propose modeling the data-generating mechanism via a density function that is invariant under group operations representing crystal and specimen symmetries. In this setting, an appropriate density function must be invariant for the equivalence class $[g] = \{ q_c * g * q_s | q_s \in Q_s,q_c \in Q_c \}$ where $*$ is a binary operator representing quaternion multiplication, $Q_c$ and $Q_s$ are groups representing the crystal and specimen symmetries of a crystal, respectively. In other words, the density function, $f$, must satisfy $f(g_j) =  f(g_k)$ for all $g_j,g_k \in [g]$. It was shown that the symmetric Bingham distribution, \begin{equation}
\text{SB}(g|\Lambda ,V) = \frac{1}{JK}\frac{1}{F(\Lambda)} \sum_{j = 1}^J\sum_{k = 1}^K\exp(-\sum_{d = 1}^4 \lambda_d (([\text{v}_d]_{j,k})^\top g)^2),
\end{equation} satisfies this desired invariance property. The symmetric Bingham distribution is parameterized by the vector $\Lambda = (\lambda_1 ,\lambda_2 ,\lambda_3 ,\lambda_4)^\top$ and the orthonormal matrix $V$, which are related to scale and location of the modes of the distribution, respectively. To ensure model identifiability, we enforce the inequality constraint $\Lambda_1 \geq \Lambda_2 \geq \Lambda_3 \geq \Lambda_4 = 0$. We let $v_d$ represent the $d^{th}$ column vector of $V$, and $[v_d]_{j,k} = q_j * v_d * q_k$. The normalizing constant $F(\cdot)$ is a hypergeometric function \citep{bingham_1974}, which is computationally intractable and is approximated using interpolation from a pre-computed look-up table. For a given material, $Q_c$ and $Q_s$ are known and depend on the material type and how it is processed.

The statistical problem is to infer the parameters of a mixture of symmetric Bingham distributions,  
\begin{equation} \label{SBM}
\text{SBM}(g|\mathbf{\Lambda},\mathbf{V}, \alpha, M) = \sum_{m = 1}^M \alpha_m \text{SB}(g|\Lambda_m ,V_m),
\end{equation} where $\mathbf{\Lambda}$ is the set of vectors $\{\Lambda_m\}_{m = 1}^M$, $\mathbf{V}$ is the set of matrices $\{V_m\}_{m = 1}^M$, and the mixture weights, $\alpha = (\alpha_1,\ldots,\alpha_M)^\top$ must satisfy $\sum_m^M \alpha_m = 1$ with $0\leq\alpha_m\leq 1$ for all $m = 1,\ldots,M$. The number of mixture components, $M$, is treated as an unknown quantity. 

We consider the following hierarchical Bayesian model,
\begin{align}
g_i|\mathbf{\Lambda},\mathbf{V}, \mathbf{\alpha}, M &\overset{iid}{\sim}
\text{SBM}(g_i|\mathbf{\Lambda},\mathbf{V}, \mathbf{\alpha}, M),\quad
i = 1,\ldots, n, \label{eq:likelihood}\\
\lambda_{d,m}|\lambda_{d-1,m} & \overset{iid}{\sim} Exponential(\mu) 1_{\lambda_{d,m}>\lambda_{d,m-1}}, \quad \lambda_{4,m} = 0, \; m = 1,\ldots, M, \; d = 1,2,3, \label{eq:prior_lambda}\\
%\nonumber\\
V_m &\overset{iid}{\sim} Uniform(O(4)),\;
m = 1,\ldots, M,\\
\alpha |M &\sim Dirichlet_M(\beta),  \\
M - 1 &\sim Poisson(\nu)1_{M\leq M_{max}} \label{eq:prior_M}.
\end{align}
The hyper-parameter $\mu$ controls the spread of the truncated exponential prior on the location variables, so that choosing large $\mu$ will result in a more diffuse prior. The truncation ensures that the support of the density follows the ordering constraint on the $\Lambda_{d,m}$s. A uniform prior on the space of orthonormal matrices, $O(4)$ is selected for $V_m$, so that no modal orientation is preferred a-priori. The Dirichlet prior for the mixture weights with hyper-parameter $\beta$ is chosen so that the prior density is close to a discrete uniform distribution on $\{1,\ldots,M\}$. In order to facilitate computation, we bound the number of mixture components, $M$, above by $M_{max}$ and below by 1, using a truncated Poisson prior. When the hyper-parameter $\nu$ is small this coincides with preferring more parsimonious models. The resulting posterior distribution has density,
\begin{equation}\label{posterior}
p(\mathbf{\Lambda},\mathbf{V},\alpha,M|\mathbf{g}) \propto p(\mathbf{g}|\mathbf{\Lambda},\mathbf{V} \alpha,M)p(\mathbf{\Lambda})p(\mathbf{V})p(\alpha|M)p(M),
\end{equation}
where we use the shorthand notation, $p(\mathbf{g}|\cdot)$, to denote the joint likelihood of the observations defined by a mixture of symmetric Bingham densities and we use $p(\cdot)$ to denote prior distributions for respective parameters, defined above in Equations \ref{eq:prior_lambda} - \ref{eq:prior_M}. Posterior inference for this model is based on samples obtained via reversible-jump Markov chain Monte Carlo \citep{green_1995}, which enables sampling on subspaces of variable dimension. Using Monte Carlo estimation methods with posterior samples, we will be able to approximate the posterior predictive distribution (PPD) for a new orientation dataset $\mathbf{g^{\text{new}}}$,
\begin{equation}\label{ppd}
p(\mathbf{g}^{\text{new}}|\mathbf{g}) =\int p(\mathbf{g}^{\text{new}}|\mathbf{\Lambda},\mathbf{V}, \alpha,M)p(\mathbf{\Lambda},\mathbf{V}, \alpha,M|\mathbf{g}) d(\mathbf{\Lambda},\mathbf{V},\alpha,M),
\end{equation} 
which allows us to interpret our model on the space of the data, and compare model fit to current kernel density estimation methods for ODF inference.

\section{Markov chain Monte Carlo Implementation}

In order to sample from the posterior distribution with density $p(\mathbf{\Lambda},\mathbf{V}, \alpha, M | \mathbf{g})$, we combine the reversible-jump Markov chain Monte Carlo (RJMCMC) technique from \cite{green_1995} to transverse dimensions of the parameter space of mixture distributions with a parallel tempering algorithm \citep{Geyer1991} employing adaptive proposal techniques \citep{andrieu_2008} to enable efficient exploration of this challenging posterior model. Throughout this section, we use $\mathbf{\Lambda}^\text{cur},\mathbf{V}^\text{cur}, \alpha^\text{cur}, M^\text{cur}$ to denote current parameter states of a Markov chain and $\mathbf{\Lambda}^\text{can},\mathbf{V}^\text{can}, \alpha^\text{can}, M^\text{can}$ to denote candidate states that are potentially attained.

To implement RJMCMC we first randomly select a dimension change from a mixture distribution with $M^\text{cur}$ components. These changes are best summarized through the transition matrix.

\begin{table}[h]
\centering $P$ = 
\begin{tabular}{llllllll}
                            & 1      & 2      & 3      & 4      & $\ldots$ & $M_{max}-1$    & $M_{max}$      \\
\multicolumn{1}{l|}{1}      & .7     & .3     & 0      & 0      & $\ldots$ & 0      & \multicolumn{1}{l|}{0}      \\
\multicolumn{1}{l|}{2}      & .15    & .7     & .15    & 0      & $\ldots$ & 0      & \multicolumn{1}{l|}{0}      \\
\multicolumn{1}{l|}{3}      & 0      & .15    & .7     & .15    & $\ldots $& 0      & \multicolumn{1}{l|}{0}      \\
\multicolumn{1}{l|}{4}      & 0      & 0      & .15    & .7     & $\ldots$ & 0      & \multicolumn{1}{l|}{0}     \\
\multicolumn{1}{l|}{$\vdots$} & $\vdots$ & $\vdots$ & $\vdots$ & $\vdots$ & $\ddots$ & $\vdots$ & \multicolumn{1}{l|}{\vdots} \\
\multicolumn{1}{l|}{$M_{max}$}      & 0      & 0      & 0      & 0      & $\ldots$ & .3     & \multicolumn{1}{l|}{.7} 
\end{tabular}
\end{table}
The elements of this transition matrix were set to strike a balance between transitioning between dimensions too seldomly or too frequently, which has worked well in the examples considered in Section \ref{EgSec}. This balance enables the algorithm to explore regions of high posterior probability within a certain number of mixture components, while considering the possibility of transitioning to a different dimensional mixture model. After proposing a candidate number of mixture components, we use deterministic mappings to map parameter values between dimensions in the following way. If $M^\text{can} <  M^\text{cur}$, we truncate the mixture component with the smallest mixture weight, and absorb the smallest mixture weight. If $M^\text{can} >  M^\text{cur}$, we keep the first $M^\text{cur}$ mixture components and add a uniform component as our $(M^\text{cur} + 1)^{th}$ component which has a modal orientation for the location parameter. This is an example of a birth-death mapping, for which the Jacobian is equal to 1, and simplifies the calculation of the acceptance probability for a change in dimension. This mapping is designed to move between regions of different dimensional subspaces with similar posterior probability, resulting in higher probability of acceptance and a more efficient sampler in terms of Monte Carlo error. The candidate mixture component model parameters remain the same as the current parameters when $M^\text{can} = M^\text{cur}$. If the candidate number of mixture components is different from the current number, this induces candidate values for remaining parameters, $\mathbf{\Lambda}^\text{can},\mathbf{V}^\text{can},\alpha^\text{can}$. We accept the dimension change with probability,
\begin{equation}\label{acceptance_M}
a_M = \min\{1,\frac{p(\mathbf{g}|\mathbf{\Lambda}^\text{can},\mathbf{V}^\text{can}, \alpha^\text{can},M^\text{can})p(\mathbf{\Lambda}^\text{can})p(\mathbf{V}^\text{can})p(\alpha^\text{can}|M^\text{can})p(M^\text{can})P_{M^\text{can},M^\text{cur}}}{p(\mathbf{g}|\mathbf{\Lambda}^\text{cur},\mathbf{V}^\text{cur}, \alpha^\text{cur},M^\text{cur})p(\mathbf{\Lambda}^\text{cur})p(\mathbf{V}^\text{cur})p(\alpha^\text{cur}|M^\text{cur})p(M^\text{cur})P_{M^\text{cur},M^\text{can}}}\}.
\end{equation}

Given a fixed number, $M^\text{cur}$, of mixture components, we use the Metropolis-Hastings algorithm to propose new values for $\mathbf{\Lambda}^\text{cur},\mathbf{V}^\text{cur}, \alpha^\text{cur}$. Using symmetric densities, we propose $\mathbf{\Lambda}^\text{can},\mathbf{V}^\text{can}, \alpha^\text{can}$ and accept the proposals with probability,

\begin{equation}\label{acceptance}
a =  \min\{1,\frac{p(\mathbf{g}|\mathbf{\Lambda}^\text{can},\mathbf{V}^\text{can}, \alpha^\text{can},M^\text{cur})p(\mathbf{\Lambda}^\text{can})p(\mathbf{V}^\text{can})p(\alpha^\text{can}|M^\text{cur})}{p(\mathbf{g}|\mathbf{\Lambda}^\text{cur},\mathbf{V}^\text{cur}, \alpha^\text{cur},M^\text{cur})p(\mathbf{\Lambda}^\text{cur})p(\mathbf{V}^\text{cur})p(\alpha^\text{cur}|M^\text{cur})}\}.
\end{equation}
We iterate these two proposal mechanisms for a large number of times, $N$, saving current states as samples from the posterior distribution of interest.

We use a simple initialization for the current parameter states. The initial parameters will correspond to a uniform symmetric Bingham distribution. i.e. $M^{[0]} = 1,\mathbf{\Lambda}^{[0]}= (0,0,0,0)',\alpha^{[0]} = 1$, and $\mathbf{V}^{[0]} = [\bar{g}_1,null(\bar{g}_1')],$ where $\bar{g}_1$ is a modal orientation. We will use modal orientations when traversing dimension, so we will compute $\bar{g}_1,\ldots,\bar{g}_{M_{max}}$ from the data prior to running the RJMCMC. In Algorithm \ref{alg:rjmcmc}, we describe the reversible-jump step and proposal mechanisms for candidate parameter values. We tune the proposal variances $b,c,d$ during the burn-in period of the RJMCMC to achieve moderate acceptance rates \citep{andrieu_2008}. In order to efficiently sample from multimodal posterior distributions and to improve mixing, we implement parallel tempering MCMC. 

Parallel tempering relies on the ability to run multiple Markov chains in parallel whose limiting distributions are related. In this case, for a set of temperatures, T indexed by $t$, we choose to run Markov chains in parallel that individually target
\begin{equation}
p(\mathbf{\Lambda},\mathbf{V}, \alpha,M|\mathbf{g})^{T(t)} \propto (p(\mathbf{g}|\mathbf{\Lambda},\mathbf{V}\alpha,M)p(\mathbf{\Lambda})p(\mathbf{V})p(\alpha|M)p(M))^{T(t)}, \quad t = 1,\ldots,|T|,
\end{equation}
where we choose $T = (1,.9,.8,.7,.6,.5,.4,.3,.2,.1)^\top$. For each of the parallel chains, within-chain state adaptations are made using RJMCMC with acceptance probabilities induced by the tempered posterior distributions. Then, between-chain swaps of the current states of each of the Markov chains are proposed and accepted according to a Metropolis-Hastings acceptance probability. The full parallel tempering algorithm details, along with expression for within- and between-state acceptance probabilities are provided in Algorithm \ref{alg:pt}. This sampling strategy works in the following way: chains targeting low temperatures rapidly traverse the parameter space exploring regions of lower posterior probability that may be difficult to attain in higher temperature chains, allowing this information to be passed along to the chain of interest through the increasing temperature scheme. The samples obtained from the chain targeting $t = 1$ are ultimately of interest as this is the posterior for which we wish to perform inference. 

Not only do these sampling techniques enable posterior inference, but they also allow prediction based on samples from the posterior predictive distribution. Typical approaches of numerical or Monte Carlo integration methods to approximate the PPD of equation \ref{ppd} are computationally intractable due to the large, complex parameter space and the normalizing constant of the symmetric Bingham distribution that is expensive to evaluate repeatedly. Instead, we propose to base an approximation of the PPD on a kernel density estimate of posterior predictive draws. We sample $n_\text{new}$ many draws, denoted $\mathbf{g}^\text{new} = g^\text{new}_1,\ldots,g^\text{new}_{n_\text{new}}$, from the PPD by using a conditional sampling method. With MCMC posterior samples stored from either Algorithm \ref{alg:rjmcmc} or \ref{alg:pt}, a draw from the posterior predictive distribution can be sampled by simulating a quaternion $g^\text{new}$ from a mixture of symmetric Bingham distributions parameterized by a randomly selected MCMC state of the posterior parameters. This process is repeated $n_\text{new}$ many times to produce independent, identically distributed draws from the PPD and is fully outlined in Algorithm \ref{alg:ppd}. A smooth PPD estimate based on these samples is produced through a kernel density estimate, as detailed in Section \ref{EgSec}.

\section{ODF Inference Examples}\label{EgSec}

We validate our methods through ODF visualization and model fit comparison to kernel density estimation methods, for which we employ the MATLAB package for material texture analysis, MTEX \citep{bachmann_2010}.

While it is possible to interpret parameters in this context, understanding the behavior of the resulting ODF is crucially important in the application area. Since the ODF support is a subset of a 4-dimensional Euclidean space, results are often mapped to lower dimensions for visualization. The correspondence between quaternions and Euler angles, also used for describing 3-dimensional rotations, can be used to plot ODF intensities in a 3-dimensional coordinate system or a pole figure can be used to visualize stereographic projections of ODF intensity at different orientations. MTEX enables both of these visualization methods. Please see \cite{bunge2013texture} and \cite{kocks1998texture} for detailed description of the construction and interpretation of these visualizations. 

For visualizations, it is common in the materials science and geological literature to normalize ODFs against a uniform distribution. This convention is maintained in the figures and visualizations presented below. Uniform texture is commonly observed for materials which solidified from the liquid state or precipitated form solution (e.g. most ceramics, and metals before any deformation processing). While this normalization can be mathematically inconvenient, it aids in interpreting the ODF. For example in Figure \ref{SFeg} below, the ODF value of 5 indicates that a particular orientation occurs will be observed 5 times more frequently than in a material with a uniform distribution.

The MTEX toolbox also implements kernel density estimation methods for ODF estimation, which are widely used in practice. The methods require the specification of a kernel $\psi_h$ which is a radially symmetric density function on quaternions that respects the the crystal and specimen symmetry groups: $Q_c$ and $Q_s$. Based on observed quaternions, $\mathbf{g}$, the kernel density estimator is defined as,
\begin{equation}\label{eq:kde}
  \widehat{p_\text{KDE}(g)}  = \sum_{i=1}^n \psi_h(g*g_i^{-1}),
\end{equation}
where a de la Vall\'{e}e Poissin kernel (\citet{schaeben1999}) with bandwidth parameter, $h$, selected through Kullback-Leibler cross validation (\citet{hall1987}) are default in MTEX.

\subsection{The Santa Fe ODF}\label{SFsec}

\begin{figure}[t!]
\centering
\includegraphics[width =\textwidth]{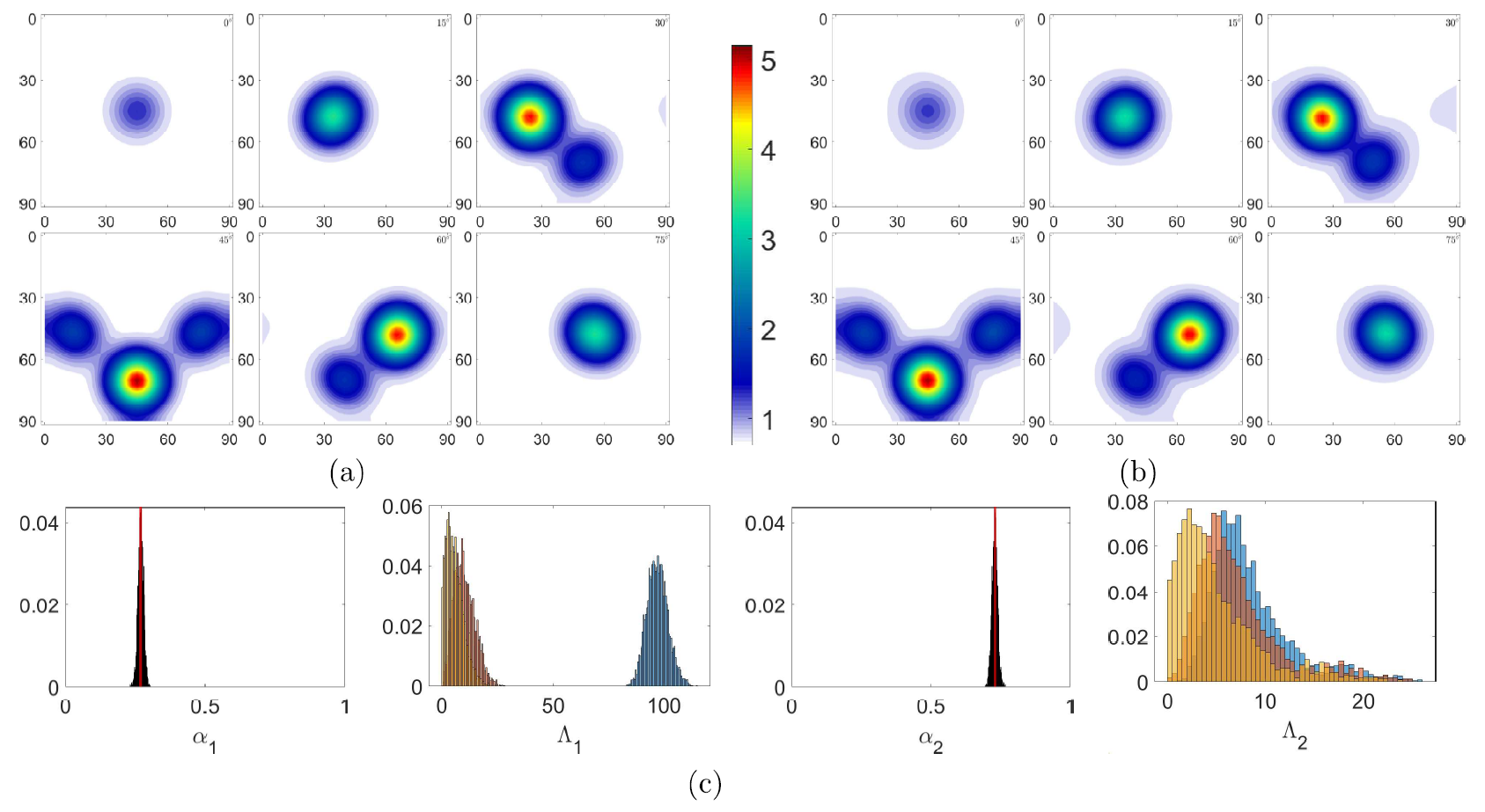}
\caption{(a) The SantaFe ODF ground truth from MTEX. (b)  %Mixture of symmetric Bingham distributions 
MAP estimate from our model. (c) Histograms of mixture weights and scale parameters for 2 mixture components. The mixture weights for the de la Vall\'{e}e Poissin kernel and a uniform components of the SantaFe ODF are shown as a vertical red line.}
\label{SFeg}
\end{figure}

The SantaFe ODF in MTEX, is a standard test case for inferring ODFs from crystallographic data. The ODF is a mixture of a de la Vall\'{e}e Poissin kernel and a uniform distribution with  0.27 and 0.73 weights respectively, and accounts for 24 crystal symmetries and 2 specimen symmetries. We observed $n = 10000$ sample orientations from this ODF, and based inference on 30000 RJMCMC iterations (Algorithm \ref{alg:rjmcmc}) with $M_{max} = 3$ dimensions. Figure \ref{SFeg}(a) and Figure \ref{SFeg}(b) compares the ground truth SantaFe ODF and a plug-in estimate of the ODF based on maximum a-posteriori (MAP) estimates of the parameter values from our model. The MAP ODF is calculated by evaluating a  symmetric Bingham mixture distribution at MAP parameter values, and represents the symmetric Bingham mixture ODF under which the data is most probable. This visualization is produced by viewing the ODF in the Euler angle coordinate system. There appears to be very little difference between the MAP estimate and the ground truth ODF. Based on Monte Carlo estimates from posterior samples, our model correctly favors a two-component mixture model with $P(M = 2|\mathbf{g}) \approx 0.9861$ and $P(M = 3|\mathbf{g}) \approx 0.0139$. Figure \ref{SFeg}(c) displays histograms of posterior sample for the mixture weights and scale parameters for each mixture component. One component in the scale parameter in the first mixture component is much larger than the others, indicating that the component is concentrated around one modal orientation and behaves similarly to the de la Vall\'{e}e Poissin kernel. The components of the scale parameter for the second component are small and close to one another, indicating that the component behaves similarly to a uniform distribution. With this interpretation of the scale parameters of the components in mind, the model provides reasonable inference for the corresponding mixture weights. Conditional on $M =2$, the model estimates reasonable mixture weights with $E[\alpha_1|\mathbf{g},M=2] \approx 0.2606$ and $E[\alpha_2|\mathbf{g},M=2] \approx .7394$, with the histograms of the mixture weights tightly concentrated around these values.

\begin{figure}[t!]
    \centering
\includegraphics[width = \textwidth]{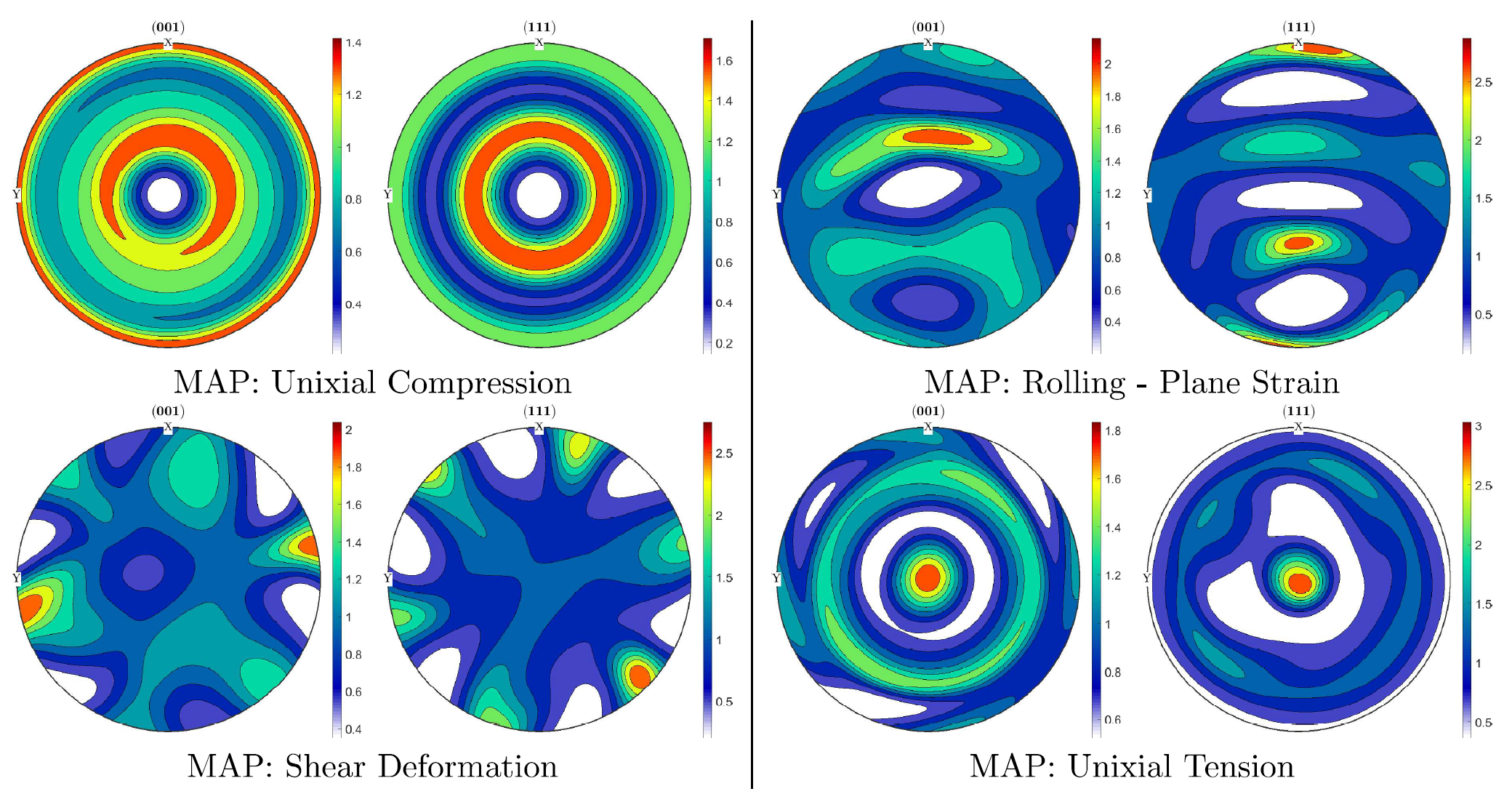}
\caption{Pole figure visualizations of MAP ODFs based on data from materials that underwent unixial compression, rolling - plane strain, shear deformation, and  unixial tension.}
\label{processedMAP}
\end{figure}

\begin{figure}[t!]
    \centering
\includegraphics[width = \textwidth]{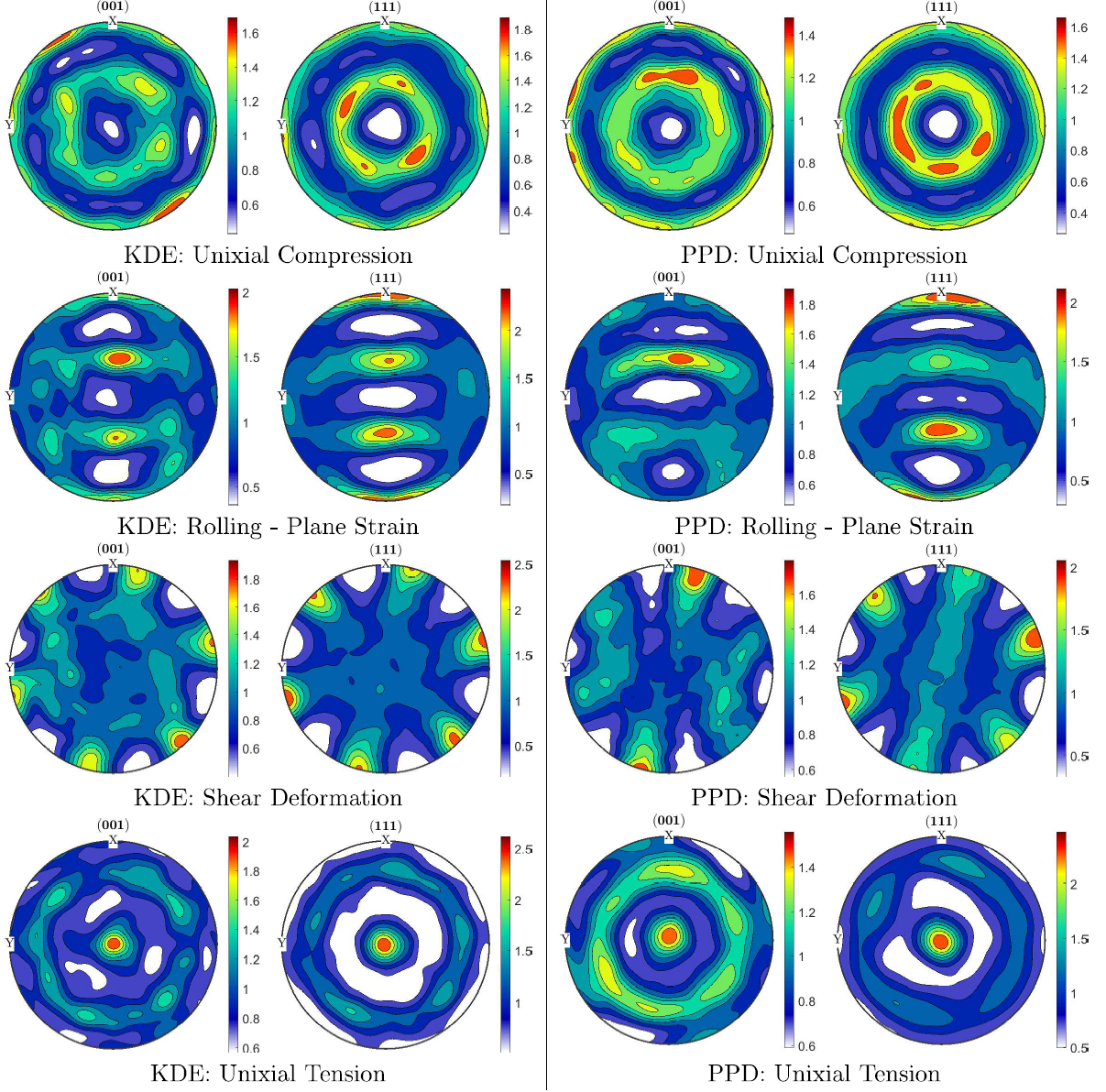}
\caption{Pole figure visualizations of kernel density estimates (left) compared to the posterior predictive distributions (right) based on sample orientations from materials that underwent unixial compression, rolling - plane strain, shear deformation, and  unixial tension (top) - (bottom).}
\label{ProcessedEg}
\end{figure}

\subsection{Processed Materials}\label{ProcessedSec}

The application of different stresses cause crystals in a material to orient themselves. In four different settings we inferred ODFs from sample orientations, $n = 2000$, of different materials with 48 crystal symmetries that underwent unixial compression, rolling - plane strain, shear deformation, and unixial tension. In each of these settings, a Taylor model was used to simulate these stresses to 20\% strain. We consider a maximum number of mixture components to be $M_{max} = 5$, while forcing one mixture component to correspond to a uniform distribution. As previously described, a uniform distribution is the expected ODF before processing. We use 10000 RJMCMC iterations (Algorithm \ref{alg:pt}) from the primary chain of parallel tempering to make inference. As in the previous example, we view a symmetric Bingham mixture distributions evaluated at estimated MAP parameters values in Figure \ref{processedMAP}. Since the true underlying ODFs that generate the different datasets is unknown, we assess model fit by viewing the posterior predictive distribution with a comparison to a kernel density estimate of an ODF based on the same observations. The PPD is used for comparison against the KDE methods since it incorporates posterior uncertainty, unlike the simple MAP estimate. As previously discussed, the PPD is approximated via kernel density estimation of posterior predictive draws outlined in Algorithm \ref{alg:ppd}. Pole figure representations of estimated ODFs for each of the stresses are contrasted in Figure \ref{ProcessedEg}. These examples demonstrate that the general patterns of the PPDs are similar to that of KDEs, while our methods achieve this model fit much more parsimoniously. In contrast with PPD estimates, the MAP estimates are unable to express important, yet less probable features into the estimated ODF, as they are only based on a single point estimate of parameter values; the difference is most notable in the rolling example.

\section{Discussion}

In this work, we have constructed a physically realistic hierarchical Bayesian model for the ODF of polycrystalline materials, performed posterior inference with the help of specialized sampling techniques, and proposed the use of PPD as an informative visualization tool in this setting. In Example \ref{SFsec}, we illustrated that the proposed model and implementation are well calibrated, and provided interpretable parameter inference with uncertainty quantification. Additionally, as shown by Example \ref{ProcessedSec}, the model is able to parsimoniously achieve similar model fit in common processed examples when compared with current kernel density estimation methods. While these results themselves are useful, the modelling framework sets the stage for additional analysis of the data through hypothesis testing and model comparison. A key open area where this work will play a critical role is the validation of computational simulations for material processing, where ODFs estimated from simulations containing only a small number of crystallites need to be quantitatively compared against experimental ODF estimates made from a much larger number of orientation measurements. Additionally, the authors wish to take full advantage of the uncertainty provided by the Bayesian approach by combining posterior inference for individual parameters with visualization of uncertainty in a predicted ODF. 

\section{Acknowledgments}
This work is partially funded by the Airforce Research Laboratory project FA8650-17-1-5277 \emph{Ensemble predictions of material behavior for ICMSE for additive structures} and by seed funding provided through the The Center for Emergent Materials an NSF MRSEC at The Ohio State University.

\bibliographystyle{chicago}
\bibliography{matuk_BIPM_references}

\appendix

\section{Algorithm details}

\begin{algorithm}[h]
\label{alg1}
\caption{Reversible-jump Markov chain Monte Carlo}\label{alg:rjmcmc}
\begin{algorithmic}[1]
\State Initialize $M^\text{cur},\mathbf{\Lambda}^\text{cur},\mathbf{V}^\text{cur},\alpha^\text{cur} = M^{[0]},\mathbf{\Lambda}^{[0]},\mathbf{V}^{[0]},\alpha^{[0]}$
\For{ \text{iter} = 1: N}
\State Propose $M^\text{can}$ through $P_{M^\text{cur},M^\text{can}}$
\If{$M^\text{can} == M^\text{cur} - 1$}
\State Find index $m$ corresponding to $\text{min}(\alpha^\text{cur})$
\State Truncate mixture component corresponding to index $m$
\State Set $\alpha^\text{can}_{p} = \alpha^\text{can}_{p} + \text{min}(\alpha^\text{cur})/M_{\text{\text{can}}}$ for $p \neq m$
\EndIf
\If{$M^\text{can} == M^\text{cur} + 1$}
\State $s = sort((L\cdot \alpha^\text{cur},unif(0,1)^\top)$ ($L$ is a lower triangular matrix of ones)
\State $\alpha^\text{can} = s - (0;s(1:M^\text{cur}))^\top$
\State Sort $\alpha^\text{can}$ so that the smallest value is in the $M^\text{can}$ component.
\State $\Lambda_{M^\text{can}} = (0,0,0,0)^\top$
\State $V_{M^\text{can}} = (\bar{g}_{M^\text{can}}, null(\bar{g}^\top_{M^\text{can}}))$
\EndIf
\State Calculate $a_M$ from Equation \ref{acceptance_M}
\If{$\text{uniform}(0,1) < a_M$}
\State Set $M^\text{cur} = M^\text{can}$ and $\mathbf{\Lambda}^\text{cur},\mathbf{V}^\text{cur},\alpha^\text{cur} = \mathbf{\Lambda}^\text{can},\mathbf{V}^\text{can},\alpha^\text{can}$
\EndIf
\State Draw $s \sim MVN(L\cdot \alpha^\text{cur},b\cdot I_{M^\text{cur}}$ and normalize so $s$ sums to $1$
\State Set $\alpha^\text{can} = s - (0;s(1:M^\text{cur}-1))^\top$
\If{$\alpha_{min} < 0$}
\State Repeat proposal of $\alpha^\text{can}$ until $\alpha^\text{can}\geq0$
\EndIf
\For{$m = 1:M^\text{cur}$}
\State Let $randVec \sim MVN((1,0,0,0)^\top,d\cdot I_4)$
\State $randRot = RandVec/norm(randVec)$
\State $v^\text{can}_{1,m} = v^\text{cur}_{1,m}*randRot$
\State $V_{m}^\text{can} = (v_{1,m}^\text{can},null((v^\text{can}_{1,m})^\top))$
\State $\lambda_{1:3,m}^\text{can} = diag( MVN(\lambda^\text{cur}_{1:3,m},c\cdot I_3)$, $\lambda_{4,m}^\text{can} = 0$
\If {$\lambda^\text{can}_{1,m}\geq \lambda^\text{can}_{2,m} \geq \lambda^\text{can}_{3,m} \geq \lambda^\text{can}_{4,m} = 0$}\State  keep $\Lambda^\text{can}_{m}$
\Else  
\State Repeat proposal of $\Lambda^\text{can}_{m}$ until conditions satisfied
\EndIf
\EndFor
\State Calculate $a$ from Equation \ref{acceptance}
\If {$\text{uniform}(0,1) < a$}
\State $\mathbf{\Lambda}^\text{cur},\mathbf{V}^\text{cur},\alpha^\text{cur} = \mathbf{\Lambda}^\text{can},\mathbf{V}^\text{can},\alpha^\text{can}$
\EndIf
\State Save $M^{[\text{iter}]},\mathbf{\Lambda}^{[\text{iter}]},\mathbf{V}^{[\text{iter}]},\alpha^{[\text{iter}]} = M^\text{cur},\mathbf{\Lambda}^\text{cur},\mathbf{V}^\text{cur},\alpha^\text{cur}$
\EndFor
\end{algorithmic}
\end{algorithm}

\begin{algorithm}[h!]
\caption{Parallel tempering}\label{alg:pt}
\begin{algorithmic}[1]
\State Initialize $M^{\text{cur},{[t]}},\mathbf{\Lambda}^{\text{cur},{[t]}},\mathbf{V}^{\text{cur},{[t]}},\alpha^{\text{cur},{[t]}} = M^{[0]},\mathbf{\Lambda}^{[0]},\mathbf{V}^{[0]},\alpha^{[0]}$ for all $t \in T$
\For{\text{iter} = 1:N}
\For{$t \in T$}
\State Obtain $M^{\text{cur},{[t]}}\Lambda^{\text{cur},{[t]}},V^{\text{cur},{[t]}},\alpha^{\text{cur},{[t]}}$ from Algorithm \ref{alg:rjmcmc} lines 2-35, with the following changes for $a_M$ and $a = a(t)$:
\State $\qquad a_M := a_M(t) = \min\{1,\big (\frac{p(\mathbf{g}|\mathbf{\Lambda}^\text{can},\mathbf{V}^\text{can}, \alpha^\text{can},M^\text{can})p(\mathbf{\Lambda}^\text{can})p(\mathbf{V}^\text{can})p(\alpha^\text{can}|M^\text{can})p(M^\text{can})P_{M^\text{can},M^\text{cur}}}{p(\mathbf{g}|\mathbf{\Lambda}^\text{cur},\mathbf{V}^\text{cur}, \alpha^\text{cur},M^\text{cur})p(\mathbf{\Lambda}^\text{cur})p(\mathbf{V}^\text{cur})p(\alpha^\text{cur}|M^\text{cur})p(M^\text{cur})P_{M^\text{cur},M^\text{can}}}\big )^{T(t)}\},$
\State $\qquad a := a(t) = \min\{1,\big (\frac{p(\mathbf{g}|\mathbf{\Lambda}^\text{can},\mathbf{V}^\text{can}, \alpha^\text{can},M^\text{cur})p(\mathbf{\Lambda}^\text{can})p(\mathbf{V}^\text{can})p(\alpha^\text{can}|M^\text{cur})}{p(\mathbf{g}|\mathbf{\Lambda}^\text{cur},\mathbf{V}^\text{cur}, \alpha^\text{cur},M^\text{cur})p(\mathbf{\Lambda}^\text{cur})p(\mathbf{V}^\text{cur})p(\alpha^\text{cur}|M^\text{cur})}\big )^{T(t)}\}.$
\EndFor
\State Randomly select $t\in (1,\ldots,|T|-1)^\top$
\State Calculate $a_{swap} = \frac{p(\mathbf{g}|\mathbf{\Lambda}^{\text{cur},{[t]}},\mathbf{V}^{\text{cur},{[t]}},\alpha^{\text{cur},{[t]}},M^{\text{cur},{[t]}})p(\mathbf{\Lambda}^{\text{cur},{[t]}})p(\mathbf{V}^{\text{cur},{[t]}})p(\alpha^{\text{cur},{[t]}}|M^{\text{cur},{[t]}})p(M^{\text{cur},{[t]}})}{p(\mathbf{g}|\mathbf{\Lambda}^{\text{cur},{[t+1]}},\mathbf{V}^{\text{cur},{[t+1]}},\alpha^{\text{cur},{[t+1]}},M^{\text{cur},{[t+1]}})p(\mathbf{\Lambda}^{\text{cur},{[t+1]}})p(\mathbf{V}^{\text{cur},{[t+1]}})p(\alpha^{\text{cur},{[t+1]}}|M^{\text{cur},{[t+1]}})p(M^{\text{cur},{[t+1]}})}$
\If {$\text{uniform}(0,1) < a_{swap}$}
\State Swap current states of chains at temperatures $t$ and $t + 1$
\EndIf
\State Save $M^{[\text{iter},t]},\mathbf{\Lambda}^{[\text{iter},t]},\mathbf{V}^{[\text{iter},t]},\alpha^{[\text{iter},t]} = M^{\text{cur},{[t]}},\mathbf{\Lambda}^{\text{cur},{[t]}},\mathbf{V}^{\text{cur},{[t]}},\alpha^{\text{cur},{[t]}}$
\EndFor
\end{algorithmic}
\end{algorithm}

\begin{algorithm}[h!]
\caption{Conditional sampling for the posterior predictive distribution}\label{alg:ppd}
\begin{algorithmic}[1]
\State Let $M^{[\text{j}]},\mathbf{\Lambda}^{[\text{j}]},\mathbf{V}^{[\text{j}]},\alpha^{[\text{j}]}, \quad j = 1,\ldots,N$ denote post burn-in RJMCMC samples from Algorithm \ref{alg:rjmcmc} or RJMCMC samples from the primary chain Algorithm of \ref{alg:pt} 
\For{\text{i} = 1:$n_\text{new}$}
\State Randomly select $j' \in \{1,\ldots,N \}$
\State Sample $g^\text{new}_i \sim SBM(g|\mathbf{\Lambda}^{[\text{j'}]},\mathbf{V}^{[\text{j'}]},\alpha^{[\text{j'}]},M^{[\text{j'}]})$
\State Save $g^\text{new}_i$ as an iid sample from the PPD (Equation \ref{ppd})
\EndFor
\end{algorithmic}
\end{algorithm}

\end{document}